\documentclass[12pt]{article}

\usepackage{scicite}
\usepackage{times}
\usepackage{graphicx}

\begin{document} 
\baselineskip24pt

\title{Observation of Microwave Shielding of Ultracold Molecules}

\author
{Lo\"ic Anderegg,$^{1,2}$ Sean Burchesky,$^{1,2}$ Yicheng Bao,$^{1,2}$\\Scarlett S. Yu,$^{1,2}$ Tijs Karman,$^{3,4}$ Eunmi Chae,$^{5}$ Kang-Kuen Ni,$^{1,2,6}$ \\Wolfgang Ketterle,$^{2,7}$ John M. Doyle$^{1,2}$
\\
\normalsize{$^{1}$Department of Physics, Harvard University, Cambridge, MA, USA}\\
\normalsize{$^{2}$Harvard-MIT Center for Ultracold Atoms, Cambridge, MA, USA}\\
\normalsize{$^{3}$ITAMP, Harvard-Smithsonian Center for Astrophysics, Cambridge, MA, USA}\\
\normalsize{$^{4}$Radboud University, Institute for Molecules and Materials,}\\
\normalsize{ Heijendaalseweg 135, 6525 AJ Nijmegen, Netherlands}\\
\normalsize{$^{5}$Department of Physics, Korea University, Seongbuk-gu, Seoul, Republic of Korea}\\
\normalsize{$^{6}$Department of Chemistry and Chemical Biology, Harvard University, Cambridge, MA, USA}\\
\normalsize{$^{7}$Department of Physics, Massachusetts Institute of Technology, Cambridge, MA, USA}\\
}

\date{\today}

\newenvironment{sciabstract}{%
\begin{quote} \bf}
{\end{quote}}

\begin{sciabstract}
Harnessing the potential wide-ranging quantum science applications of molecules will require control of their interactions. Here, we use microwave radiation to directly engineer and tune the interaction potentials between ultracold calcium monofluoride (CaF) molecules. By merging two optical tweezers, each containing a single molecule, we probe collisions in three dimensions. The correct combination of microwave frequency and power creates an effective repulsive shield, which suppresses the inelastic loss rate by a factor of six, in agreement with theoretical calculations. The demonstrated microwave shielding shows a general route to the creation of long-lived, dense samples of ultracold molecules and evaporative cooling.

\end{sciabstract}

\maketitle

Applications of ultracold molecules in quantum simulation, precision measurement, ultracold chemistry, and quantum computation~\cite{Carr2009review, Bohn2017}, have led to rapid progress in direct cooling ~\cite{shuman10, barry14, truppe17,anderegg17, collopy18}, assembly~\cite{Ni2008KRb, Takekoshi2014RbCs, Park2015NaK, Gregory2016RbCs, Guo2017NaRb, Rvachov2017LiNa, Liu2018, Cairncross2021, Marco2019}, trapping~\cite{anderegg18, williams18, McCarron18}, and control of molecules~\cite{Ospelkaus2010hfcontrol, Park2017NaK, Seeselberg2018rotcoh, Chou2017}. Engineering and control of molecular interactions will enable or enhance many of these appliactions. In particular, collisional interactions play a critical role in the ability to cool and, therefore, control molecules. While there has been some success with sympathetic and evaporative cooling ~\cite{KRbevap, Son2020},  these efforts have been hindered by large inelastic loss rates for both reactive and non-reactive molecular species in optical traps ~\cite{Ye2018, Gregory2019RbCsSticky, Hu2019KRbReaction, Cheuk2020}. Suppressing these inelastic losses, and more generally tuning interactions, is key to effective evaporative cooling of ultracold molecules and quantum applications.

A path to this is shielding, whereby molecules can be repelled from short range distances where inelastic processes occur. Various shielding schemes for atoms~\cite{Bali1994,Suominen1995,Julienne1996,Napolitano1997} and molecules have been proposed\cite{Gorshkov2008shield, Quemener2016shieldE, Gonzalez-Martinez2017adimtheory, Karman2018mwshield, Karman2019, Karman2020}. Recently, a scheme using DC electric fields to generate repulsive interactions was demonstrated in a 2D geometry for KRb molecules~\cite{Matsuda2020}.
Here, we report microwave shielding of  $^{40}$Ca$^{19}$F molecules in three dimensions using optical tweezer traps. By tuning the microwave frequency from blue to red detuned, the system goes from shielded to ``anti-shielded", changing the inelastic collision rate by a factor of 24, in agreement with theoretical calculations.

\begin{figure}[t]
\centering
\includegraphics[width=.8\textwidth]{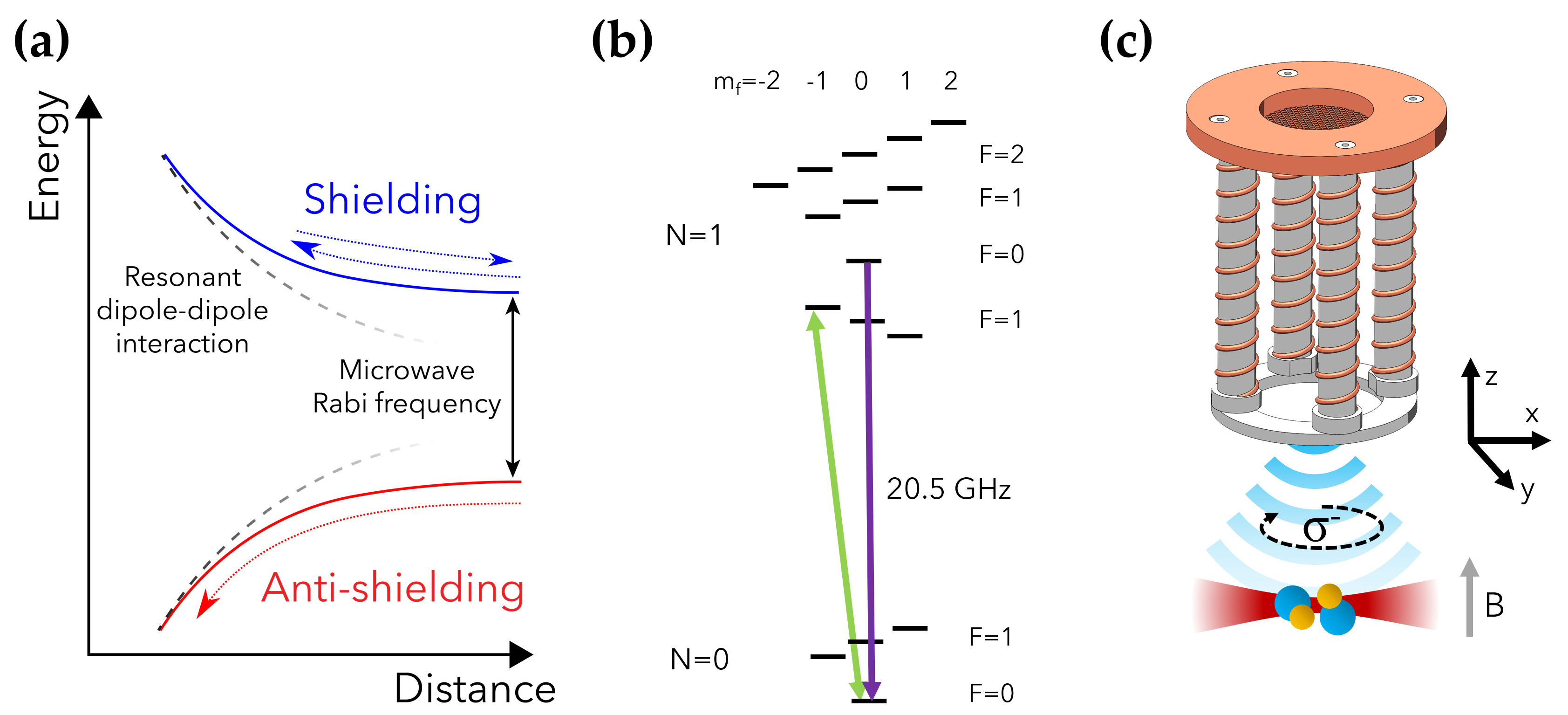}
\caption{\label{fig:MWspec}\textbf{Microwave shielding overview} (a) Diagram of the shielding process. The upper dressed state leads to a repulsive potential, preventing the molecules from reaching short range and undergoing loss. (b) CaF energy levels of the $X^2 \Sigma^+$ electronic ground state showing the N=1 and N=0 rotational states separated by 20.5~GHz. The shielding transition is shown in green. The purple arrow shows the Landau-Zener sweep to the absolute ground state.  (c) Experimental schematic showing the relative orientations of the helical antenna with respect to the tweezers. The tweezer light is linearly polarized along the z axis, parallel to the magnetic field, B.}
\end{figure}

The microwave shielding mechanism studied here works as follows. Continuous, near resonant microwave fields dress the molecular states, generating an oscillating dipole moment in the CaF molecule that gives rise to strong, long-range dipolar interactions~\cite{Yan2020}. With the correct microwave frequency, this interaction is repulsive. Additionally, the dipolar interaction significantly enhances the elastic collision rate resulting in a high elastic-to-inelastic collisional ratio, which is a key feature for evaporative cooling. In the microwave shielding scheme we use, the uppermost dressed state adiabatically converts to the repulsive branch of the dipole-dipole interaction. This leads to a classical turning point at long range \cite{Karman2018mwshield, Karman2019}, as shown in Fig. \ref{fig:MWspec}(a), preventing molecules from reaching short range where they would be lost with high probability\cite{Cheuk2020}.
There will be residual inelastic loss at long range, predicted to be a result of non-adiabatic transitions (so-called ``microwave-induced loss") \cite{Karman2018mwshield, Karman2019}.
Coupled-channel calculations have shown that effective shielding requires circular polarization and high Rabi frequencies of the microwave field~\cite{Karman2018mwshield, Karman2019}.
Circular polarization provides coupling to the repulsive branch of the resonant dipole-dipole interaction regardless of the orientation of the collision axis relative to the molecule orientation, resulting in shielding in the bulk, i.e. three dimensions. Rabi frequencies can be made high enough to create a large gap between field-dressed levels, ensuring adiabaticity during the collision.

Our experiment starts from a magneto-optical trap (MOT) of $^{40}$Ca$^{19}$F molecules~\cite{anderegg17}. We use $\Lambda$-enhanced grey molasses cooling to load molecules into a conservative 1064 nm optical dipole trap~\cite{anderegg18,Cheuk18}.  Single molecules are then transferred into two 780 nm optical tweezer traps \cite{Anderegg2019tweezer}. The tweezers have a beam waist of about $1.6~ \mu$m and a depth of  $1.8$~mK. Light assisted collisions caused by the $\Lambda$-cooling light during tweezer loading ensures no more than a single molecule in each tweezer. As detailed previously~\cite{Cheuk2020} the two tweezer traps can be merged to create a colliding pair of molecules in a single trap. This is accomplished using a single 780~nm laser source to create one stationary trap, and using an acousto-optical deflector (AOD), one steerable trap. The light for theses two traps is combined, then focused down, forming two tightly focused tweezer traps that can be merged. We measure a typical molecule temperature of $\sim$96~$\mu$K \cite{submat}, both before and after merging. The molecules occupy many spatial modes therefore the collisions are three-dimensional in nature.

Once the tweezers are loaded, we apply an optical pumping pulse to populate the $|N=1,J=1/2, F=0, m_f=0 \rangle$ state (Fig.  \ref{fig:MWspec}(b)). Next, we use a Landau-Zener microwave sweep to move the population to the absolute ground state ${|N=0,J=1/2,F=0,m_f=0 \rangle}$. This transition is nominally dipole forbidden but an applied 4 Gauss magnetic field mixes in the ${|N=0,F=1,m_f=0 \rangle}$ state, providing a significant transition dipole moment. To remove any remaining population in the N=1 rotational level, we apply a 5~ms pulse of resonant light, heating the N=1 molecules out of the tweezer trap.  The two molecules, both in the ground internal state, are then merged together into a single tweezer for collisions to take place. At this point, the shielding is turned on for a variable amount of time before the tweezers are separated, and the molecules are transferred back to the N=1 manifold for imaging.

\begin{figure}[t]
\centering
\includegraphics[width=.8\textwidth]{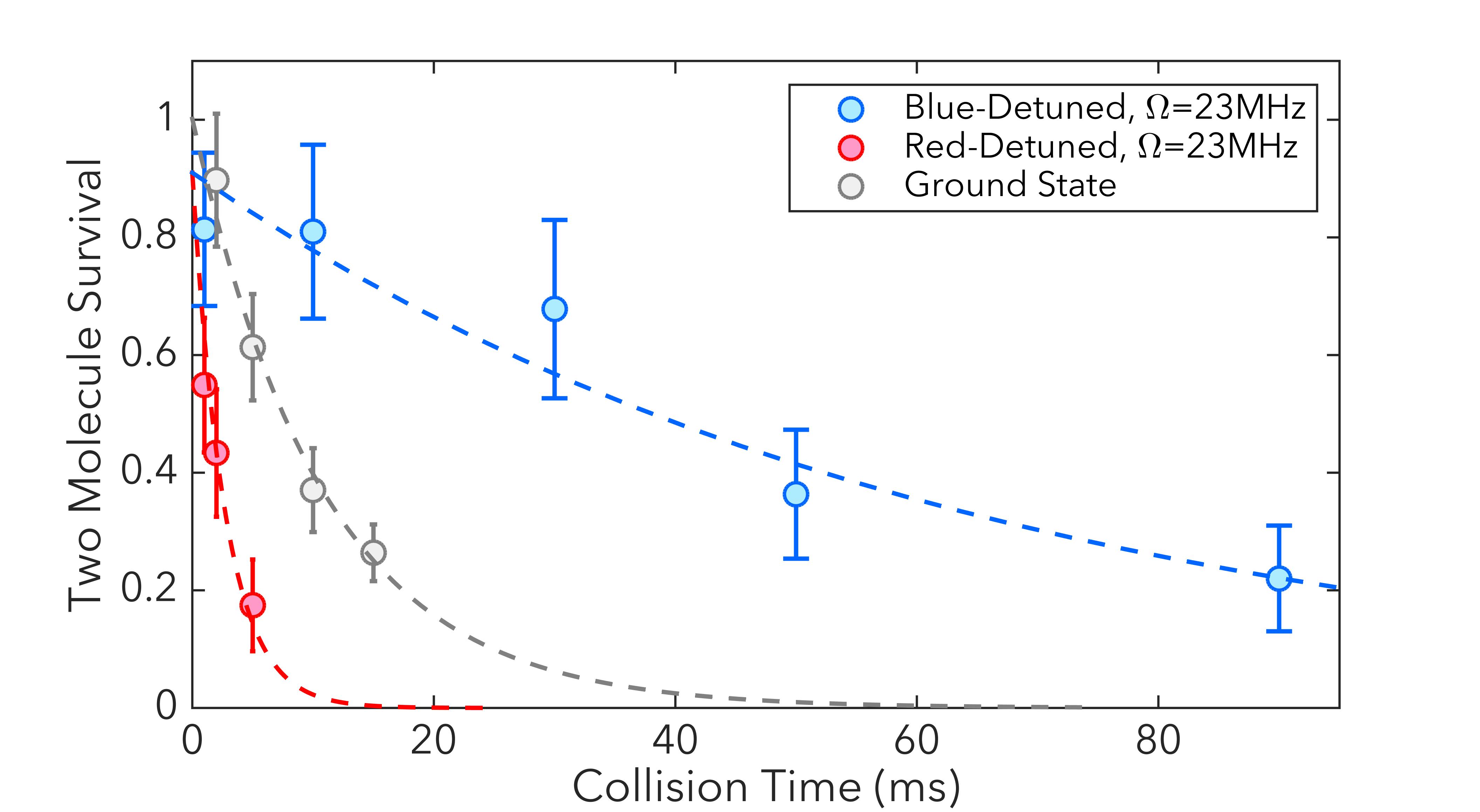}
\caption{\textbf{Microwave shielding of CaF collisions} The grey trace (10.8~ms) shows the bare two body loss of unshielded ground state collisions. The blue trace (64~ms) shows the shielded loss rate at a Rabi frequency of 23 MHz, and magnetic field of 27 G while blue detuned.  The red trace (2.7~ms)  shows the loss rate while red detuned with a Rabi frequency of 20 MHz, and magnetic field of 27 G.}
\label{collisiondata}
\end{figure}

Circularly polarized microwaves are generated by a $2\times2$ helical antenna array~\cite{Kraft1996}, Fig. \ref{fig:MWspec}(c). The helical antennas are designed for axial mode operation, creating circular polarization with a helicity set by the winding of the antenna. An array is used to increase the cleanliness of the circular polarization and the overall output power. The 20.5~GHz microwaves are generated from mixing a low phase noise 18.5~GHz source with a 2~GHz source locked to a low noise oscillator \cite{submat}. The 20.5~GHz signal is then amplified and split into four paths of equal length through phase stable cables. Each antenna has a separate 5~W microwave amplifier and a mechanical phase shifter. The microwaves propagate into a stainless-steel vacuum chamber through a glass window along the z-axis, defined as the direction of the magnetic field, Fig. \ref{fig:MWspec}(c). We determine the polarization of the microwave field by measuring the Rabi frequency of the $\sigma^{+}$, $\sigma^{-}$, and $\pi$ transitions between the states ${| N=1,J=1/2,F=0,m_f=0 \rangle}$ and ${| N=0,J=1/2,F=1, m_f=\pm{1},0 \rangle}$. Accounting for the magnetic field dependent matrix elements, the $\sigma^{+}$ and $\sigma^{-}$ field components indicate the degree of circular polarization in the plane transverse to the axial magnetic field, while the $\pi$ component of the field is related to the tilt angle of the polarization ellipse relative to the z-axis. Using the measured Rabi frequencies, we then adjust the phases of the four individual antennas to maximize the  target circular field component, while minimizing the other two polarizations. The helical antenna array generates clean circular polarization in free space, however the reflections from metal components in and around the vacuum chamber degrade the polarization cleanliness to a power ratio of right- to left-handed circular polarization of 100 \cite{submat}. 

\begin{figure}[t]
\centering
\includegraphics[width=.8\textwidth]{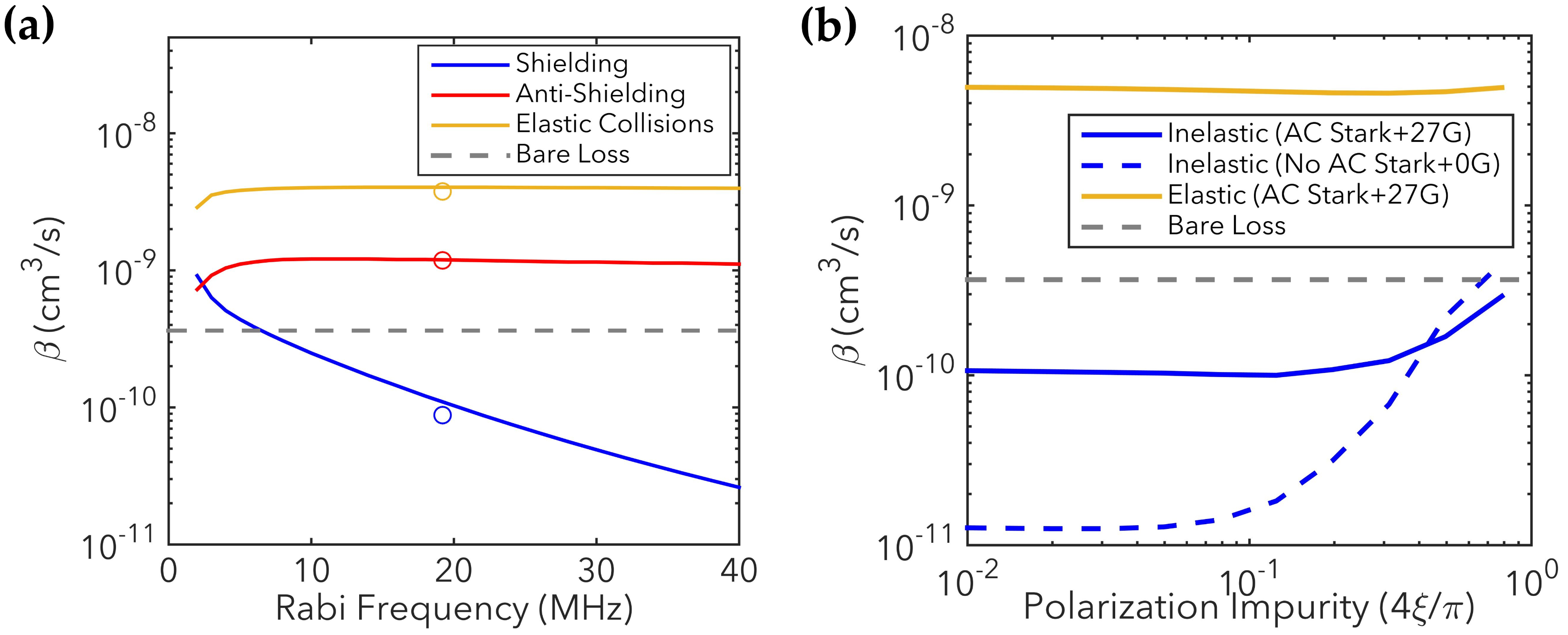}
\caption{\label{fig:theory}\textbf{Theory}
(a) Rate coefficient versus Rabi frequency. Both shielding (blue) and anti-shielding (red) are shown. The elastic rate is shown in yellow. The solid lines are results without including spin, the circles include spin.
(b) Plot of loss versus microwave ellipticity angle, $\xi$\cite{submat}, for a Rabi frequency of 23 MHz. The tweezer trap's tensor AC Stark shift is the dominant factor reducing the effective degree of shielding. The elastic rate is nearly unaffected by the AC Stark shift and magnetic field.
}
\end{figure}

To create collisional shielding, we use the ${|N=1,J=1/2,F=1,m_f \rangle}$ hyperfine manifold, with an applied 27 Gauss magnetic field. The magnetic field direction is such that the upper state in the manifold ${|N=1,J=1/2,F=1,m_f=-1  \rangle}$ is driven by the high purity circular polarization. After merging the tweezers, we prepare the upper dressed state by switching on low power microwaves with a frequency a few MHz blue detuned of $|N=1,J=1/2,F=1,m_f=-1  \rangle$. Then, adiabatically, the amplitude of the microwaves is ramped to full power in $\sim$100 $\mu$s.  Using the highest energy $m_f$ level ensures that the upper dressed state does not cross any other levels as the microwave power is ramped up. The lifetime of the single-particle dressed state in the optical tweezer is limited by the phase noise of the microwave source to $>500$ ms.

The collisional lifetime of the bare ground state is the reference for our shielding performance comparison. We measure the trap frequency and temperature of the bare ground state molecules to be the same as molecules prepared in the upper dressed state, thus ensuring that the density of microwave dressed molecules and bare ground state molecules are comparable. The ratio of the measured lifetimes is thus the ratio of the 2-body loss rates.

At a Rabi frequency of 23~MHz, and dressing 3 MHz blue detuned from the $|N=0,F=0,m_f=0\rangle$ to $|N=1,J=1/2,F=1,m_f=-1 \rangle$ transition in a 27 Gauss field, as in Fig. \ref{collisiondata}, the shielded lifetime was 64 ms ($\beta=7.2(2.0) \times 10^{-11}$ cm$^3$/s), six times longer than the bare ground state lifetime of 10.8 ms  ($\beta=4.2 (0.8) \times 10^{-10}$ cm$^3$/s). The ratio of the lifetimes are in agreement with a coupled channel loss rate calculation, Fig. \ref{fig:theory}. We find experimentally that the shielded lifetime is relatively independent of the polarization purity from $100:1$ to $10:1$ in power at a Rabi frequency of $\sim23$ MHz.

While the upper dressed state produces a repulsive shielding potential, the lower dressed state adiabatically connects to the attractive branch of the dipole-dipole interaction as the molecules approach during the collision, causing anti-shielding ~\cite{Karman2018mwshield, Karman2019}. Guided by this theory, we prepare the lower dressed state by flipping the direction of the magnetic field, which effectively swaps the handedness of the microwaves such that the lowest $m_f$ level ( $|N=1,J=1/2,F=1,m_f=+1 \rangle$) is now the one being driven most strongly by the circularly polarized microwaves. We prepare the lower dressed state with a microwave power ramp, with the microwaves $7$ MHz red detuned. We measure this anti-shielded lifetime to be $2.7$ ms ($\beta=1.7 (0.5) \times 10^{-9}$ cm$^3$/s), or a factor of about four faster than the bare ground state and a factor of twenty-four faster than the shielded state, see Fig. \ref{collisiondata}.

We use coupled-channel methods to calculate microwave shielding of CaF molecules.
Similar to previous work\cite{Karman2019}, the colliding molecules are modeled as rigid rotors interacting through dipole-dipole interactions and with external magnetic and microwave fields.
For details see \cite{submat}.
In contrast to previous studies, we include the tensor AC Stark shift caused by the intense tweezer light.
At short range, a fully absorbing boundary condition is imposed that yields universal loss in the absence of microwave dressing. Non-adiabatic transitions between dressed states lead to microwave-induced loss,
whereby the microwave Rabi frequency is converted into kinetic energy.
Short-range losses occur to a lesser extent as the potentials involved are mainly repulsive.
The results of two sets of calculations are shown in Fig.~\ref{fig:theory}.
First, shown in Fig.~\ref{fig:theory}(a), we assume the microwave polarization is perfectly circular about the magnetic field and tweezer polarization directions.
Cylindrical symmetry is exploited to expedite the calculations.
Second, shown in Fig.~\ref{fig:theory}(b), the ellipticity of the microwave polarization is added, breaking cylindrical symmetry, which makes the computations more demanding such that explicitly accounting for (hyper)fine structure becomes intractable.
Hence, we make the approximation of treating spin implicitly by an enhanced rotational $g$-factor and test the accuracy of this approximation in Fig.~\ref{fig:theory}(a), see \cite{submat} for a detailed discussion.

Previous theoretical work on microwave shielding \cite{Karman2018mwshield, Karman2019} indicated a strong dependence on the polarization. The tensor Stark shift due to the optical tweezer light aligns the molecules which competes with the resonant dipolar interactions that lead to shielding.
The coupled channel calculations performed here indicate this limits shielding for perfectly circular polarization but reduces the sensitivity to polarization imperfections, see Fig. \ref{fig:theory}(b). The $|N=1,J=1/2,F=1,m_f \rangle$ hyperfine manifold, used for shielding, has a significant tensor polarizability \cite{submat}. The tweezer in which the collisions take place is linearly polarized along the z-axis, parallel to the magnetic field and perpendicular to the plane containing the polarization ellipse of the microwaves. At a tweezer trap depth of 1.8 mK for the ground state and no applied magnetic field, we observe a splitting of 10 MHz between the $m_f=0$ and $m_f=\pm{1}$ states. This tensor Stark shift is the dominant limiting factor of the observed shielding process.

\begin{figure}[t]
\centering
\includegraphics[width=.8\textwidth]{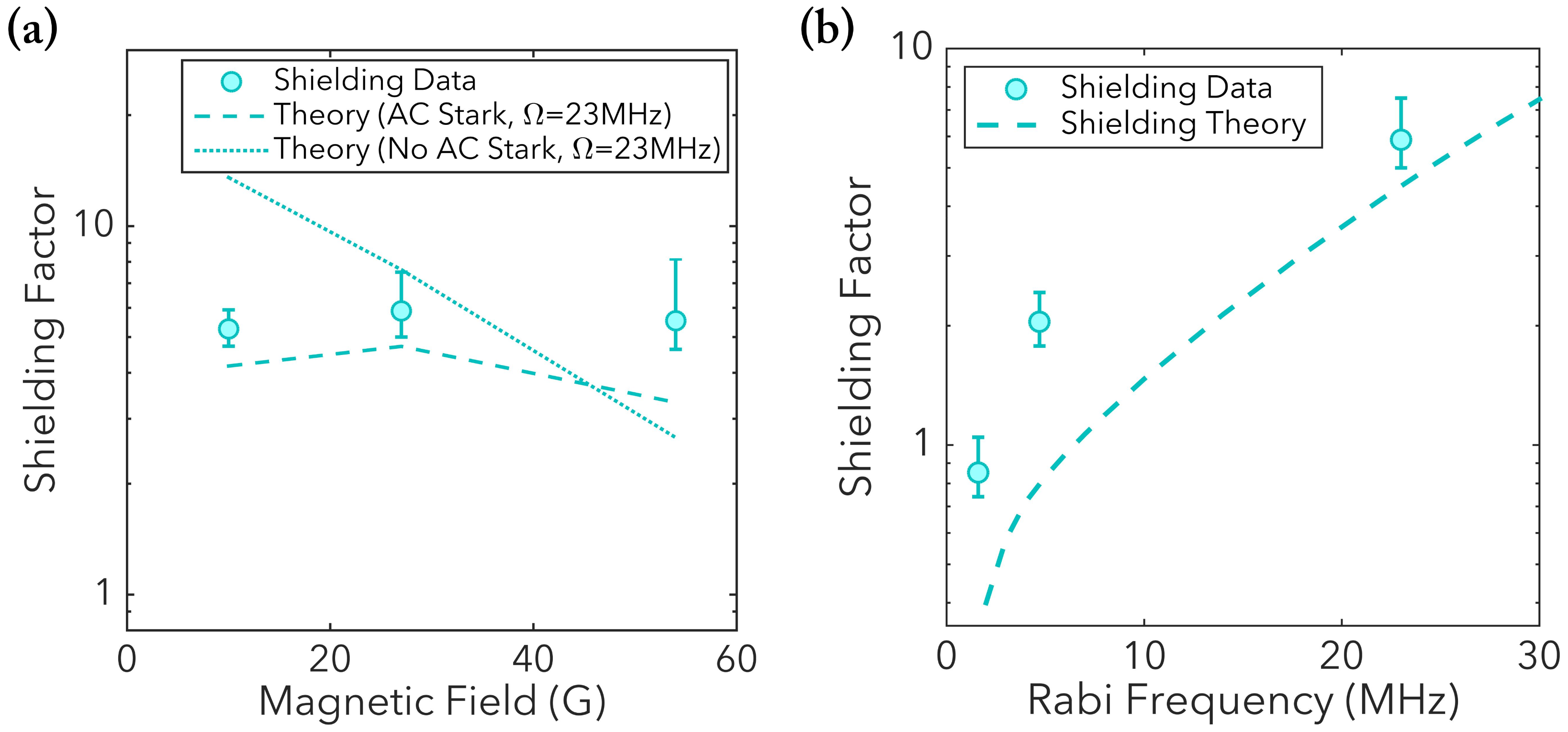}
\caption{\textbf{Dependence of shielding on microwave power and magnetic field} The shielding factor is the ratio of the bare loss rate to the measured loss rate. (a) Shielding factor versus magnetic field at a Rabi frequency of 23 MHz. We find the effect of shielding to be robust over this range of magnetic fields, a result of the tensor AC Stark shift from the trap light. (b) Shielding factor versus Rabi frequency. We find the crossover point where shielding begins to be around 3 MHz. The experimental data is taken at 27 G while the theory curve does not include the effect of spin.}
\label{scans}
\end{figure}

It has been shown previously \cite{Karman2018mwshield} that the CaF($^2\Sigma$) fine structure can limit the effectiveness of microwave shielding,
leading to losses enhanced by orders of magnitude compared to calculations neglecting (hyper)fine structure or to $^1\Sigma$ bialkali molecules, with much weaker hyperfine interactions.
Here, we use the ${|N=0,J=1/2,F=0  \rangle}  \rightarrow {|N=1,J=1/2,F=1  \rangle}$ transition, see Fig.~\ref{fig:MWspec}(a), to achieve shielding.
At low magnetic fields, the electron and nuclear spins in these states are approximately coupled to a total spin singlet, which effectively eliminates the fine structure couplings.
As shown in Fig.~\ref{scans}(a), our calculations predict an enhancement of the shielding at low magnetic field but only in the absence of tensor AC Stark shifts.
Including tensor AC Stark shifts, this interaction dominates the loss at low magnetic field,
and the interplay between these two effects results in only a weak magnetic field dependence,
see \cite{submat} for a detailed discussion.
This is in agreement with experimentally observed shielding lifetimes that are similar for 10 Gauss, 27 Gauss, and 54 Gauss, Fig. \ref{scans}(a).

As predicted from theory, the shielding effect shows a clear power dependence, Fig. \ref{scans}(b). The loss rate increases from $\beta=7.2(2.0)\times 10^{-11}$ cm$^3$/s  to $\beta=2.1 (0.5) \times 10^{-10}$ cm$^3$/s when the microwave power is reduced from $20$ to $5$ MHz of Rabi frequency on the $\sigma^{-}$ transition, while keeping the polarization unchanged. Decreasing the power further to $1.5$ MHz of Rabi frequency, we measure an increased loss rate of $\beta=5.0 (1.1) \times 10^{-10}$ cm$^3$/s, slightly higher than the bare ground state. The losses measured are almost entirely from long-range microwave driven non-adiabatic transitions, therefore the bare ground state loss rate is not a lower bound ~\cite{Karman2018mwshield, Karman2019}. As the gap between the upper dressed state providing the repulsive potential and the lower dressed state decreases, the loss rate at low Rabi frequency is faster than the bare ground state loss rate.

In conclusion, we demonstrate microwave shielding of inelastic collisions in three dimensions with ultracold CaF molecules in an optical tweezer trap. The relative ratios of experimentally measured 2-body lifetimes agree well with results of coupled channel calculations and the qualitative features of shielding theory. By blue detuning the microwaves, we observe a factor of six suppression of inelastic loss in the shielded upper dressed state, relative to the bare ground state. By red detuning, we create an anti-shielded lower dressed state, leading to an enhanced loss rate. This shielding mechanism may be extended to a wide range of molecules, including polyatomic molecules~\cite{kozyryev16, Mitra2020}. It is notable that the predicted elastic scattering rate with microwave dressing is greatly enhanced, leading to a ratio ($\gamma$) of elastic to inelastic rates of over $50$, more than enough for effective direct evaporative cooling. Theory also suggests that the shielding is limited by AC stark shifts from the tweezer traps, indicating the possibility of improved shielding by using lower optical trap intensities.

\phantom{\cite{Chen2012}}
\phantom{\cite{Liesbeth2013}}
\phantom{\cite{Childs1984}}
\phantom{\cite{Aldegunde2018}}

\bibliography{shieldbib} 
\bibliographystyle{Science}

\textbf{Funding:} This work was supported by the ARO, DOE, and NSF. LA acknowledges support from the Harvard Quantum Initiative. SB and SY acknowledge support from the NSF GRFP. EC is supported by NRF of Korea.

\end{document}